%% file: mapreduce_lamia.tex
\documentclass[10pt, conference, compsocconf]{IEEEtran}
\usepackage{nameref,amssymb}
\usepackage{subfig}
\ifCLASSINFOpdf
  \usepackage[pdftex]{graphicx}
\else
  \usepackage[dvips]{graphicx}
\fi
\usepackage[cmex10]{amsmath}
\usepackage{url}
\begin{document}
%
\title{Using MapReduce for Large--scale Medical Image Analysis}
%
\author{\IEEEauthorblockN{Dimitrios Markonis, Roger Schaer, Ivan Eggel, Henning M\"{u}ller, Adrien Depeursinge}
\IEEEauthorblockA{
University of Applied Sciences Western Switzerland (HES--SO), Business Information Systems,\\
TechnoArk 3, 3960 Sierre, Switzerland\\
Email: roger.schaer@hevs.ch}
}
\maketitle
\begin{abstract}
The growth of the amount of medical image data produced on a daily basis in modern hospitals forces the adaptation of traditional medical image analysis and indexing approaches towards scalable solutions. 
The number of images and their dimensionality increased dramatically during the past 20 years. 
We propose solutions for large--scale medical image analysis based on parallel computing and algorithm optimization. 
The MapReduce framework is used to speed up and make possible three large--scale medical image processing use--cases: 
(i) parameter optimization for lung texture segmentation using support vector machines, 
(ii) content--based medical image indexing, and 
(iii) three--dimensional directional wavelet analysis for solid texture classification. 
A cluster of heterogeneous computing nodes was set up in our institution using Hadoop allowing for a maximum of 42 concurrent map tasks. 
The majority of the machines used are desktop computers that are also used for regular office work. 
The cluster showed to be minimally invasive and  stable. 
The runtimes of each of the three use--case have been significantly reduced when compared to a sequential execution. 
Hadoop provides an easy--to--employ framework for data analysis tasks that scales well for many tasks but requires optimization for specific tasks.
\end{abstract}
\begin{IEEEkeywords}
large--scale; medical; image analysis; big data; scalability; MapReduce; Hadoop; support vector machines; content--based image retrieval; texture analysis;
\end{IEEEkeywords}
\IEEEpeerreviewmaketitle
\input{parts/1_intro}

\input{parts/2_state_of_the_art}

\input{parts/3_1_methods_overview_hadoop}

\input{parts/3_2_methods_svm}

\input{parts/3_3_methods_indexing}

\input{parts/3_4_methods_riesz3d}
\input{parts/4_1_results_global}

\input{parts/4_2_results_svm}

\input{parts/4_3_results_indexing}

\input{parts/4_4_results_riesz3d}

\input{parts/5_discussion}


%
\section*{Acknowledgments}
This work was supported by the Swiss National Science Foundation (grant 205321--130046), The HES--SO LaMIA (Large--scale Medical image Analysis) project and the EU in the context of Khresmoi (257528).
\bibliographystyle{unsrt}
\bibliography{references}

\end{document}

%% file: parts/1_intro.tex
\section{Introduction}
In the past 20 years, the amount of imaging data produced on a daily basis by companies, institutions and hospitals has grown exponentially.
With the increased use of three-- or four--dimensional imaging techniques (digital security, videos, 3D temperature maps, functional medical imaging) 
over the past few years this growth has even accelerated.
Analyzing such vast and varied amounts of data to find relevant information is becoming increasingly challenging. 
Innovative software solutions are therefore needed to achieve efficient digital visual data management, including automated data analysis and retrieval.
Recent advances in computer vision have brought promising tools for efficient management of visual data~\cite{GuR1997} in two major areas:
\begin{itemize}
\item automated analysis and event--detection,
\item content--based indexation and retrieval.
\end{itemize}
The former thoroughly analyzes visual data in order to detect, characterize and quantify anomalies or events, while the latter aims to retrieve images based on visual similarity (i.e., content--based image retrieval, or "CBIR").
Both configurations rely on the concept of visual features (e.g., color intensity, texture and shape of objects, ...) which describe the visual content.
Based on a visual feature space, the detection--based tools assign a given input instance (i.e., image or image region) to a predefined class (e.g. "normal" or "abnormal"), whereas CBIR--based approaches determine a similarity score between two instances/documents from which a ranked list of results can be built.
As a consequence, the efficiency of the system (being either detection-- or CBIR--based) is closely related to the ability to catch subtle specificities of the data.
Recent progress in image processing and machine learning makes it possible to extract flexible image features, which are then mapped to a given set of classes corresponding to a specific  task.
Such techniques allow for analyzing multidimensional images with high levels of semantics, provided that a sufficient amount of training data representing intra-class variability is available.
However, the process of extracting intricate features from large datasets of 3D/4D images, as well as training machine learning algorithms and global system optimization are extremely demanding in terms of computation time, storage capacity and network bandwidth~\cite{CHH2005}.
For instance, 3D convolutions with multi--oriented, multi--scale filter banks --- that have proven to efficiently analyze 2D --- are very computationally intensive and yield many more coefficients than the number of voxels in the original image, which can lead to memory and storage problems. The next step, which involves selecting the relevant coefficients and using them as the input for machine learning algorithms, is again highly computationally intensive, and an exhaustive search for optimal parameters requires multiple runs of global experiments.

The three key factors (computation, storage and network bandwidth) therefore represent the biggest potential bottlenecks to large--scale data analysis. 
Further studies are required to identify non--optimized components and data analysis processes and most importantly propose flexible and scalable infrastructures that are able to cope with the exponential growth of visual data. Several approaches exist:
\begin{itemize}
\item Single Host (server with 12 physical processor cores with hyper--threading\footnote{Hyper-threading: technology from Intel\textsuperscript{\textregistered} which makes each physical processor core appear to the Operation System as 2 virtual cores, which improves the execution of multi--threaded code. See \texttt{http://en.wikipedia.org/wiki/Hyper\-threading} (as of 29 August 2012) for more information.}  \& 96 GB of RAM (Random Access Memory)),
\item Small local cluster (8 hosts with 2--4 physical cores with hyper--threading and 16 GB of RAM, plus the above--mentioned server),
\item Alternative infrastructures, such as GPUs (Graphical Processing Units),
\item Cloud computing infrastructures, such as Amazon's Elastic Cloud Compute (EC2).
\end{itemize}
In the context of this paper, the first two solutions are tested and compared, in order to determine the best possible option for a given analysis task.
Finding the optimal solution is a problem of potential interest to many companies and also research projects.


%% file: parts/2_state_of_the_art.tex
\section{Related work}
Due to the growth in the data size and the development of new computationally intensive algorithms, research has been performed in creating parallel processing algorithms~\cite{GeO1993} and developing cloud computing systems~\cite{RCL2009,CRB2011}. 
MapReduce~\cite{DeG2008} proposed initially by Google\footnote{\texttt{http://research.google.com/archive/mapreduce.html}, as of 29 August 2012.} has become one of the most popular distributed computing frameworks, due to its simplicity in setup and programming. 
Hadoop~\cite{Whi2010} is a very popular open source implementation of MapReduce with a large community of users. 
Although being a powerful computational tool, MapReduce should not be seen as a ``one--fits--all" solution~\cite{LLC2011}. 
For example, as demonstrated in~\cite{LNK2009} on a data warehousing use--case using astrophysical data sets MapReduce  is outperformed by database management systems (DBMS). 
In~\cite{SAD2010}, it is stated that MapReduce should be seen as an ``extract--transform--load'' (ETL) tool 
and complement DBMS in tasks that require both data warehousing and intensive processing. 

In the field of image processing, apart from cloud computing, parallelization using graphical processing units (GPU)~\cite{PSL2011} is often used. 
However, as stressed in~\cite{PSL2011}, GPU hardware architectures differ much form other architectures and should be taken into consideration when designing parallel processing algorithms. MapReduce has recently been used for large--scale image annotation~\cite{ALL2011,LCH2009} and efficient image description and analysis~\cite{WYL2010}. 
In~\cite{ALL2011} a parallel support vector machine (SVM)~\cite{Vap1995} algorithm is proposed for automatic image annotation, while in~\cite{LCH2009} a dataset of 30 million
images are automatically labeled using the MapReduce framework, although no details on the parallel implementation are given. 
Several image analysis algorithms (e.g., image feature extraction, local descriptors clustering and image registration) adaptation to the MapReduce framework is discussed in~\cite{WYL2010}.
Content--based image retrieval is another field that combines large amounts of data with computationally intensive tasks. 
For these reasons, indexing large image datasets using visual features is expected to be a well--suited task for the MapReduce framework. 
However, a few studies have also used MapReduce for the online part of the retrieval~\cite{YKA2009,LuC2008,ZLL2010}.
To overcome its inherent limitations, data warehousing and storage tools like Hive~\cite{TSJ2009} and HBase (an implementation of Google's BigTable abstraction~\cite{CDG2008}) 
that are built on top of Hadoop are often used. 
In~\cite{YKA2009} a CBIR system, NIR, on top of Hadoop, Nutch~\cite{KCS2004} and LIRe (Lucene Image Retrieval)~\cite{LuC2008} is presented. 
However, a very small data set is used for evaluation of the retrieval time. 
Another system called Distributed Image Retrieval System (DIRS) is described in~\cite{ZLL2010} uses LIRe and HBase. 
Data sets of up to 100,000 images are used for testing the query times. 
When using datasets above 20,000 images, the retrieval times reported are restrictive for online use even though they are faster than without Hadoop use.
Other approaches to deal with these challenges have included implementing efficient indexing schemes such as an inverted index~\cite{LOY2008} or locality sensitive hash
(LSH)~\cite{SMS2010} on top of the Hadoop file system (HDFS)~\cite{SKR2010}. 
While~\cite{SMS2010} reports promising results our belief is in accordance with~\cite{LLC2011,SAD2010} that for online tasks parallel DBMS should be the first choice.

In the medical field, cloud computing is also starting to find use. 
CCMedII~\cite{HYY2011} is a proposed medical information file exchanging and sharing system built on top of Hadoop. 
Medical Image File Access System (MIFAS)~\cite{YCC2010} is an access system for images using HDFS. 
MapReduce has also been used in fluorecence image analysis tasks in~\cite{ZSA2010} and as a framework of the Hadoop--GIS query system in~\cite{AjW2012} for analytical pathology imaging. These systems use HBase and Hive for data storage and warehousing respectively. 
Moreover, Hadoop has been used for anatomical landmark detection~\cite{SRZ2012} and medical image registration~\cite{KPF2009}.
%

In this paper, we use the MapReduce framework to speed up and make possible three large--scale medical image processing use--cases:
\begin{itemize}
\item parameter optimization for SVMs, which aims at identifying parameter values yielding best classification accuracy of lung textures in high--resolution computed tomography (HRCT)~\cite{DFV2011},
\item SIFT (Scale--Invariant Feature Transform) feature extraction~\cite{Low2004} and bag--of--visual--words indexing of large image datasets, which extracts features from each image and then indexes it using these features,
\item 3D texture feature extraction using the Riesz transform.
\end{itemize}

%% file: parts/3_1_methods_overview_hadoop.tex
\section{Methods}
A parallel computing environment was set up in our network using Hadoop\footnote{\texttt{http://hadoop.apache.org/}, as of 29 August 2012.}, an open--source implementation of Google's MapReduce framework. Detailed descriptions of MapReduce and Hadoop are provided in Sections~\ref{sec:MapReduce} and~\ref{sec:Hadoop}. 
The three "real--world" applications mentioned above were converted to Hadoop programs, allowing performance testing and comparison as well as identification of bottlenecks and other problems. 
The specificities of these three use cases that can potentially benefit from the MapReduce parallel computing environment are detailed in Sections~\ref{sec:methodSVM},~\ref{sec:methodImageIndexing} and~\ref{sec:methodRiesz3D}, respectively.
\subsection{MapReduce}\label{sec:MapReduce}
MapReduce is a programming model and an associated implementation developed by Google for processing large datasets. 
Typically the computation runs in parallel on a cluster of machines of up to several thousand nodes (usually commodity personal computers) in order to finish the processing task in a reasonable amount of time. 
Users define the required computation, which must be embarrassingly parallel, in terms of a map and a reduce function with the underlying system automatically distributing it across the cluster. 
The system itself manages machine failures and inter--machine communication to ensure efficiency of the network and disks. 
This approach is a reaction to code complexity by hiding the messy details of fault tolerance, data distribution and load balancing in a library. 
The computation can as well be run across multiple cores of the same machine(s)~\cite{DeG2008,Whi2010}.
Google's implementation of MapReduce runs on top of the Google File System (GFS), a scalable distributed file system for large data--intensive--applications providing fault tolerance. 
Files on GFS are typically split into chunks of 64 MB and distributed across chunkservers with a default replication factor of three~\cite{GGL2003}. 
MapReduce cannot solve every problem, but being a general data--processing tool there is a wide range of algorithms that can be expressed such as machine learning algorithms, 
graph--based problems and image analysis~\cite{Whi2010}.
\subsubsection{Programming Model}
A typical MapReduce program is split into a Map phase and into a Reduce phase. 
The Map function has a key/value pair as input and produces a set of intermediate key/value pairs as output:
\begin{equation}
 (k_1, v_1)\rightarrow\text{list}(k_2,v_2).
\end{equation}
After the map phase the MapReduce library groups all intermediate values for the same intermediate key I. 
The Reduce function accepts an intermediate key I with its set of values from the map output (supplied via an iterator) as input 
and merges these values together to produce a possibly smaller set of values as output. 
Normally the number of output values per reduce invocation is zero or one:
\begin{equation}
 (k_2, list(v_2))\rightarrow\text{list}(v_2).
\end{equation}
It is worth mentioning that the map input keys and values are related to a different domain than the keys and values of the intermediate and reduce output~\cite{DeG2008}.
\subsection{Hadoop}\label{sec:Hadoop}
%
Hadoop was created by Doug Cutting, the creator of Apache Lucene\footnote{\texttt{http://lucene.apache.org/}, as of 29 August 2012.}. 
The origins of Hadoop are found in Nutch\footnote{\texttt{http://nutch.apache.org/}, as of 29 August 2012.} (Lucene subproject), an open source web search engine supposed to scale to billions of pages. 
However, realizing that it was not possible with their architecture at that time, 
Cutting and his partner Mike Cafarella, inspired by the publication of the GFS paper in 2003~\cite{GGL2003},
decided to write an open source implementation named Nutch Distributed File System (NDFS)\footnote{\texttt{http://wiki.apache.org/nutch/\\NutchDistributedFileSystem}, as of 29 August 2012.}.
In early 2005, after the publication of the Google paper that introduced MapReduce to the world in 2004~\cite{DeG2004}, the Nutch developers presented a working implementation of MapReduce. 
As NDFS and the MapReduce implementation in Nutch were considered as potentially useful to a broader field of application
they were moved out of Nutch and became an independent subproject of Lucene called Hadoop. 
Shortly later, Cutting joined Yahoo!\footnote{\texttt{http://www.yahoo.com/}, as of 29 August 2012.}, which provided a dedicated team just for the extension of Hadoop. 
This makes Yahoo! the largest contributor of Hadoop. 
Confirmed by its success, Hadoop turned into an own top--level project at Apache in 2008~\cite{Whi2010}. 
Since then, various large companies such as Amazon\footnote{\texttt{http://www.amazon.com/}, as of 29 August 2012.}, Facebook\footnote{\texttt{http://www.facebook.com/}, as of 29 August 2012.}, Microsoft\footnote{\texttt{http://www.microsoft.com/}, as of 29 August 2012.} have started using Hadoop~\cite{HadoopPoweredBy25052012}.
%

The Apache Hadoop Common library is written in Java and consists of two main components: 
the MapReduce framework and HDFS\footnote{\texttt{http://hadoop.apache.org/hdfs/}, as of 29 August 2012.}, 
which implements a single--writer, multiple reader model~\cite{SKR2010,HadoopDocs25052012}. 
However, Hadoop does not solely support HDFS as an underlying file system. 
It also provides a general--purpose file system abstraction making it possible to integrate other storage systems, such as Amazon S3\footnote{\texttt{http://aws.amazon.com/s3/}, as of 29 August 2012.}, 
which targets the Amazon Elastic Compute Cloud\footnote{\texttt{http://aws.amazon.com/ec2/}, as of 29 August 2012.} server--on--demand infrastructure. 
In our own Hadoop environment, we exclusively make use of HDFS as file system. 
Currently, the Linux operating system is the only officially supported Hadoop production platform~\cite{Whi2010,HadoopWikipedia25052012}.

The purpose of HDFS is to store large datasets reliably and to stream them at high bandwidth to user applications. 
HDFS has two types of nodes in the schema of a master--worker pattern: a namenode, the master and an arbitrary number of datanodes, the workers~\cite{Whi2010}. 
The HDFS namespace is a hierarchy of files and directories with associated metadata represented on the namenode. 
The actual file content is split into blocks of typically 64MB where each block is typically replicated on three namenodes. 
The namenode keeps track of the namespace tree and the mapping of file blocks to datanodes. An HDFS client wanting to read a file has to contact the namenode for the locations of data blocks and then reads the blocks from the closest datanode since HDFS considers short distance between nodes as higher bandwidth between them. 
In order to keep track of the distances between datanodes HDFS supports rack--awareness. 
As soon as a datanode registers with the namenode, the namenode runs a user--configured script to decide which rack (network switch) the node belongs to. 
Rack--awareness also allows HDFS to have a block placement policy that provides a trade--off between minimizing write cost and maximizing data reliability, availability and aggregate read bandwidth. 
For the creation of a new block, HDFS places the first replica on the datanode hosting the writer and the second and third replicas on two different datanodes located in a different rack~\cite{SKR2010}.

A Hadoop MapReduce job, a unit of work that the client wants to be performed, consists of the input data (located on the HDFS), the MapReduce program and configuration information. 
Native Hadoop MapReduce programs are written in Java, however Hadoop also provides the Hadoop Streaming API which allows writing map and reduce functions in languages other than Java by using Unix standard streams as the interface between Hadoop and the user program. 
In Hadoop there are two types of nodes that control the job execution process: one job tracker, and an arbitrary number of task trackers. 
The job tracker coordinates a job run on the system by dividing it into smaller tasks to run them on different task trackers, which in turn transmit reports to the job tracker. 
In case a task fails, the job tracker is able to automatically reschedule the task on a different available task tracker. 
In order to have a task tracker run a map task the input data needs to be split into fixed--size pieces. 
Hadoop runs one map task for each split with the user-defined map function processing each record in the split. 
As soon as a map task is accomplished its intermediary output is written to the local disk. After that the map output of each map task is processed by the user--defined reduce function on the reducer. 
The number of map tasks running in parallel on one node is user--configurable and heavily dependent on the capability of the machine itself, 
whereas the number of reduce tasks is specified independently and is therefore not regulated by the size of the input. 
In case there are multiple reducers, one partition per reducer is created from the map output. 
Depending on the task to accomplish, it is as well possible to have zero reduce tasks in case no reduction is desired~\cite{Whi2010}.

%% file: parts/3_2_methods_svm.tex
\subsection{Support Vector Machines}\label{sec:methodSVM}
Segmentation and classification in medical image analysis consist of assigning a label (e.g., healthy, diseased) to a given voxel represented in a feature space 
(e.g., color intensity, texture measures). 
To automate the decision process, supervised machine learning algorithms can, after a training phase, be used to predict test voxel classes based on input visual features. 
SVMs are such a supervised learning algorithm, which allows establishing nonlinear decision boundaries between labeled instances in feature spaces. 
It aims at minimizing the generalization error, which is the classification error obtained with the instances that were not used in the training phase~\cite{Vap1995}. SVMs implement a nonlinear numerical approach to build a maximal separating hyperplane $\boldsymbol{w}$ considering a two--class problem. 
Two parallel hyperplanes are constructed symmetrically on each side of the hyperplane that separates the data. 
The goal is to maximize the distance between the two external hyperplanes, called the margin~\cite{Bur1998,Vap1995}. 
An assumption is made that the larger the margin, the better the generalization error of the classifier. Training $l_1$--norm SVMs consists in iteratively solving:
\begin{equation}\label{eq:SVM}
\min_{\boldsymbol{w},\xi,b}\left\{\frac{||\boldsymbol{w}||_1^2}{2}+C\sum_{i=1}^{n}\xi_i\right\}, 
\end{equation}
\begin{equation*}
\text{subject to}\quad y_i(K(\boldsymbol{w},\boldsymbol{x_i}) -b)\geq1-\xi_i \quad\text{and} \quad\xi_i\geq 0,
\end{equation*}
where $\xi$ is the slack variable of the soft margin, $C$ is the cost of the errors, $\boldsymbol{x_i}$ are the image instances (e.g., regions of interest to be categorized) $i=1\dots n$ represented in the feature space and $y_i$ are the corresponding labels. 
$K(\boldsymbol{x_i},\boldsymbol{x_j})$ is called the kernel function and allows a nonlinear mapping of the initial feature space. 
In this work, we consider the kernel Gaussian radial basis function as:
\begin{equation}\label{eq:gaussKernel}
K(\boldsymbol{x_i},\boldsymbol{x_j})=\exp(\frac{-||x_i-x_j||_1^2}{2\sigma^2}).
\end{equation}
To achieve optimal classification performance on the test set (i.e., generalization performance), the parameters $C$ and $\sigma$ require optimization 
using cross--validations and exhaustive grid--searches~\cite{HCL2003}, which are computationally intensive. 
In this work, we address the optimization problem by parallelizing each independent value couple $(C,\sigma)$ of the grid search.

%% file: parts/3_3_methods_indexing.tex
\subsection{Image Indexing}\label{sec:methodImageIndexing}
Content--based image search engines use one or more example images as query and search in an image data set to return a set of images that are relevant judging  the 
visual appearance represented by visual features. 
For this task, the dataset should be indexed in a comparable way beforehand. 
This indexing consists of extracting visual features from each image, creating an optimal representation of the image using these features 
and storing this representation using an efficient indexing structure. 
This pipeline, depending on visual features, indexing techniques, and size of the image dataset, can be computationally intensive.

A state--of--the--art approach in large--scale image retrieval is the ``bag--of--visual--words" representation~\cite{SiZ2003}. 
The method uses local descriptors (e.g., SIFT) to describe potentially interesting (or salient) regions of the image. 
These descriptors are quantized using a fixed set of ``visual words" $v_i$ and the 
image is described as a histogram of the occurrence frequency of the visual words. 
The final image descriptor of image $\mathbf{I}$, called bag--of--visual--words, is defined as a vector $\mathbf{F}(\mathbf{x}) = \{\bar{v}_1,\dots,\bar{v}_{k}\}$ of frequencies $\bar{v}_i$ of visual words such that, 
for each SIFT vector  $\mathbf{f}(\mathbf{x})$ extracted from the image $\mathbf{I}$:
$$\bar{v}_i=\sum_{l=1}^{n_f}\sum_{j=1}^{n_f} g_j(\mathbf{f}(\mathbf{x}_l)),\hspace{5pt} \forall i = 1,\dots,k,$$ where
\begin{equation}
 g_j(\mathbf{f}(\mathbf{x}))=
  \left\lbrace
  \begin{array}{l}
     1 \ \text{if}\ d_{\varepsilon}(\mathbf{f}(\mathbf{x}),v_j)\leq d_{\varepsilon}(\mathbf{f}(\mathbf{x}),v_l)\ \forall v_l\in \mathbf{V},\\
     0 \ \text{otherwise}. \\
  \end{array}
  \right.
\end{equation}
$d_{\varepsilon}$ refers to the Euclidean distance between two vectors. $\mathbf{V}$ is the set of visual words that is derived by clustering a training set of local descriptors in the feature space ($\mathbb{R}^{128}$ in SIFT case)
and taking the cluster centers:
\begin{equation}
 \mathbf{V}=\{ v_1,\dots,v_{k}\}, \hspace{5pt} v_i\in \mathbb{R}^{128},\hspace{5pt} i = 1,\dots,k
\end{equation}
This method, although achieving good performance, can be data set--dependent and has a variety of parameters to be tuned and optimized. 
When large--scale datasets are involved in scientific research or real--life applications, the optimization of parameters, 
the indexing of new datasets or the regular update of existing index can be highly time--consuming.

%% file: parts/3_4_methods_riesz3d.tex
\subsection{Solid 3D Texture Analysis Using the Riesz Transform}\label{sec:methodRiesz3D}
Volumetric medical image interpretation is time--consuming and error--prone 
because the radiologists have to exhaustively browse image series having sometimes several thousand slices~\cite{AWK2011}. 
Evaluation and characterization of organs and biomedical tissue require mental reconstructions of volumetric shapes and textures, which is little intuitive.
Consequently, algorithms for the characterization of solid three--dimensional texture patterns $V_{x,y,z}$ defined for each coordinate $x,y,z\in V_{x,y,z}\subset\mathbb{R}^3$ with an intrinsic dimension of 3 are becoming increasingly important to characterize and quantify the appearance of 3D medical textures in computed tomography (CT), magnetic resonance imaging (MRI), 3D ultrasound (US), and other volumetric imaging modalities used in clinical routine~\cite{KoP2000}. 
A recently developed texture analysis technique based on Riesz wavelets for 2D lung texture classification in HRCT showed promising results in~\cite{DFV2011,DFV2012a}. 
Riesz wavelets yield a multiscale and multi--orientation steerable filterbank allowing to analyze local orientations and scale with infinitesimal angular and spectral precision~\cite{UCV2011}. 
The framework is also available in 3D and was used by Chenouard et al. for biomedical image denoising in~\cite{ChU2011}. 
We use it in this work for solid 3D texture analysis using a publicly available database of synthetic textures~\cite{PMR2009}. 
The goal is to carry out preliminary studies on synthetic 3D data for a further use with volumetric medical data.

The number $M$ of Riesz components constituting the filterbank is related to the number of scales $s$, the order of the Riesz transform $N$, and the data dimensionality $d$ as~\cite{UCV2011}:
\begin{equation}\label{eq:rieszComp}
M=s\cdot\binom{N+d-1}{d-1}=s\cdot\frac{(N+d-1)!}{N!(d-1)!}.
\end{equation}
The computational complexity is therefore strongly dependent on $d$, and quickly becomes not affordable with $d=3$ on a regular desktop computer for a large image collection. 
In this work, we simply distributed the image database over the various nodes of the Hadoop cluster in subgroups of 10 volumes, from a total of 750 image series of dimensions $64\times 64\times 64$. In total, 75 map tasks were generated with this process.

%% file: parts/4_1_results_global.tex
\section{Results}
In this section, the parallel computing environment based on Hadoop is first described and evaluated using a simple ``Word Count" application as baseline.
Then, solutions are proposed for the various large--scale medical image analysis challenges introduced in Sections~\ref{sec:methodSVM},~\ref{sec:methodImageIndexing} and~\ref{sec:methodRiesz3D}.
\subsection{Hadoop Cluster}
\begin{figure*}
\centering
\includegraphics[scale=0.4]{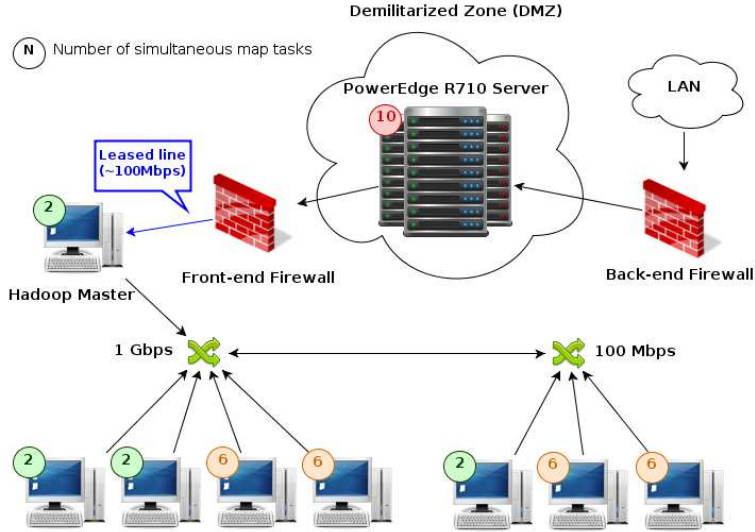}
\caption{Overview of the architecture of the Hadoop cluster. The numbers included near the machine icons indicate the number of simultaneous tasks that can be executed on each node.}
\label{fig:HadoopCluster}
\end{figure*}

The in--house Hadoop cluster that was set up for testing purposes uses a variety of nodes with a varying number of CPU (Central Processing Unit) cores and storage capacities, 
which are connected to network equipment with varying speeds (1Gbps switch, 100Mbps switch, leased line). 
The cluster is composed entirely of computers that are employed by users for their daily work and has no dedicated nodes. 
The server referenced in Fig.~\ref{fig:HadoopCluster} also hosts several other services, including several web applications and a numerical computing environment 
that can put a strain on the machine's global performance.

The computer labeled Hadoop Master in Fig.~\ref{fig:HadoopCluster} has multiple roles. It acts as:
\begin{itemize}
\item job tracker,
\item name node,
\item secondary name node (performing periodic checkpoints of the HDFS)
\footnote{\texttt{http://wiki.apache.org/hadoop/FAQ\#HDFS}, as of 29 August 2012.},
\item data node,
\item task tracker.
\end{itemize}
The master node represents a single point of failure (SPOF), which is due to the way Hadoop is structured\footnote{\texttt{http://wiki.apache.org/hadoop/SPOF}, as of 29 August 2012.}.

The remaining nodes only possess the data node and task tracker roles.
The number of simultaneous map tasks is tailored to each computer's performance. 
Since the computers are actively used during the day, at least 2 logical cores were not allocated to the Hadoop TaskTracker process, ensuring that common daily tasks could still be run smoothly. 
A larger amount of cores (14) were left unused on the server, since it is often running other processes with a heavy CPU load.

The network setup is also quite heterogeneous, with some nodes being connected to a Gigabit Ethernet switch, 
others to a slower 100Mbps switch and the server being located in a different building, which is linked to the cluster via a  leased line.
This leased line is under heavy load most of the time and its average transfer rate is close to 100Mbps.

Initial testing was done using a simple Word Count application (which counts the number of instances of each word in a text document), in order to:
\begin{itemize}
\item confirm that the cluster was functioning properly,
\item learn how to write programs for the MapReduce framework,
\item tweak the configuration of the nodes to optimize the cluster's performance.
\end{itemize}
\subsection{Job Execution Time Variability}
\begin{figure}[h]
\includegraphics[scale=0.45]{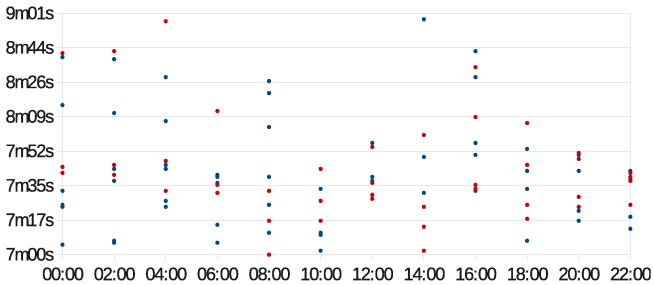}
\caption{Job execution time variability measured at several times of the day with the Word Count example. 
Blue dots indicates weekdays, whereas red dots were run on week--ends.}
\label{fig:JobVariability}
\end{figure}

In order to establish a baseline for the variability of the execution time of a job in different scenarios (during the day, at night, on week--ends, etc.), 
a basic Word Count program was run at 2--hour intervals over a period of 10 days. 
This experiment gave us a clearer picture of the impact that hardware--related factors (CPU load of the nodes, available network bandwidth, etc.) can have on the global performance of the cluster. 
Job execution time variability is depicted in Fig.~\ref{fig:JobVariability}.

%% file: parts/4_2_results_svm.tex
\subsection{Support Vector Machines}
The search for optimal SVM parameters $C$ and $\sigma$ in Eq.~(\ref{eq:SVM}) and~(\ref{eq:gaussKernel}) was translated to a MapReduce problem. 
The goal is to find the optimal value couple $(C,\sigma)$ allowing best classification performance of five lung texture patterns associated with interstitial lung diseases in high--resolution computed tomography. 
The image instances $\boldsymbol{x_i}$ are 2D $32\times32$ blocks represented in terms of energies of Riesz wavelet coefficients~\cite{DFV2011,DFV2012a}. 
The global classification accuracy is estimated using a leave--one--patient--out cross--validation (LOPO CV) with 85 cases.
First, a files containing all  possible  $C$ and $\sigma$ parameter combinations was generated. 
The latter serves as the input for the Hadoop job, where each line containing a value couple represents an independent map task. 

\begin{figure}
\centering
\includegraphics[scale=0.4]{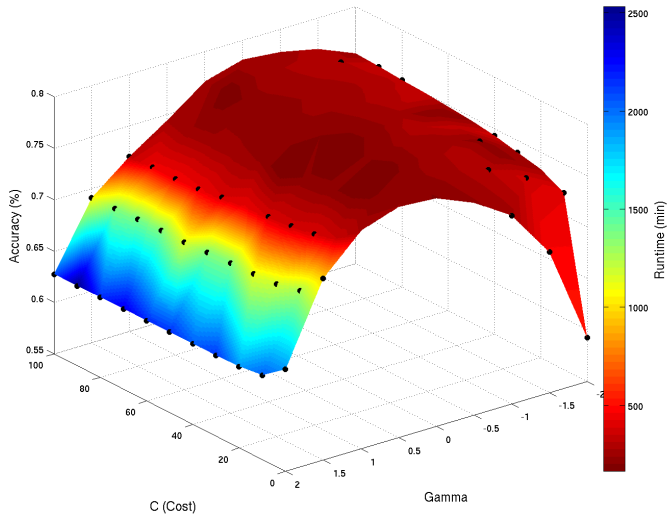}
\caption{Lung texture classification using SVMs: the relation between the runtime (color--coded) of a map task and its accuracy are depicted. Black dots indicate tasks that would be stopped by the termination clause (see Eq.~(\ref{eq:terminationClause})).}
\label{fig:SVMOVAGraph}
\end{figure}

Job execution times varied considerably depending on the values of the parameters, depending on the number of operations required to solve Eq.~(\ref{eq:SVM}). 
This was problematic since the total execution time was set by longest map tasks. 
The cause of this problem was therefore investigated and a solution to optimize the global execution time of the job was designed 
based on the assumption that the number of operations required to solve~(\ref{eq:SVM}) is linked to classification accuracy. 
$(C,\sigma)$ values that are requiring a large number of operations suggest that both the kernelization of the feature space and the cost of errors 
lead to complex decision boundaries requiring heavy optimization and result in poor classification accuracy. 
Interrupting such map tasks early is therefore a convenient way to reduce the global runtime of the Hadoop job without compromising optimal classification accuracies. 
Two approaches were considered:
\begin{itemize}
\item Limiting the runtime of a task: by using thread interruption techniques 
it would be possible to stop the execution of  a map task after a given amount of time, resulting in the loss of the calculations that were in progress. 
\item Interrupting tasks based on the reference execution time  $t_{\text{ref}}$ for solving Eq.~(\ref{eq:SVM}) taken by the fastest task after 2 folds of the LOPO CV. 
All tasks with runtimes $t_i$ exceeding $t_{\text{ref}}$ by more than a factor of $F$ after 2 LOPO CV folds are terminated and assigned an accuracy of -1 to make the interrupted tasks recognizable in the final output as:
\begin{equation}\label{eq:terminationClause}
t_i\geq F \cdot t_{\text{ref}}.
\end{equation}
\end{itemize}
The second approach was considered being a cleaner way of reducing the execution time because it does not force the interruption of the task. 
It was used with a kill factor $F=1.7$, which divided the total runtime by almost $5.5$ ($\sim$50h =\textgreater{} $\sim$9h15min), with no impact on the maximum accuracy. 
Figure~\ref{fig:SVMOVAGraph} shows the results of a job where no tasks have been interrupted,
 with superimposed black circles for each task that would be stopped when using the optimized algorithm. 
%
%

%% file: parts/4_3_results_indexing.tex
\subsection{Image Indexing}
Three subsets of the ImageCLEF\footnote{\texttt{http://www.imageclef.org/}, as of 29 August 2012.} 2011 medical task dataset~\cite{KMB2011}, containing 1,000, 10,000 and 100,000 images respectively, were used 
as test datasets. 
Two approaches were used; 1) a component--based approach and 2) a monolithic approach. 
The first one consists of two components; the feature extractor 
and the bag--of--visual--words indexer. 
In this approach 1), the intermediate output of the extractor is stored on the disk before the indexer uses it as an input. 
This is a common practice for component--based evaluation and parameter tuning. 
The monolithic approach 2) is not storing any intermediate output and performs the whole pipeline each time it is initiated.

We consider the component--based approach as being an input--output (IO)--intensive task due to the writing and reading of the outputs of the extractor component. 
Runs using multi--threading techniques with 1, 2 and 3 threads were performed on a single machine to serve as baseline. 
Preliminary results showed that even for the smaller datasets of 1,000 and 10,000 images, the multi--threading run using 3 threads outperformed the MapReduce tasks in the 42--task cluster, taking about half the time for the extractor phase. 
For 100,000 images, the multithreaded runs scale linearly for 1--3 threads. 
It requires approximately 7 hours with 3 threads for the whole pipeline (4.9 hours for the feature extractor and 2.7 hours for the bag--of--visual--words indexer). 
In comparison, MapReduce required more that 14 hours using 42 concurrent map tasks (13 hours for the feature extractor and 1.5 hours for the bag--of--visual--words indexer).
Although the MapReduce framework reached performance comparable to the single machine by increasing the number of reducers, 
the amount of resources used to achieve such a performance showed that MapReduce may not be suited for IO--intensive tasks.

\begin{figure*}
\centering
\subfloat[1,000 images]{\label{fig:1k}\includegraphics[scale=0.4]{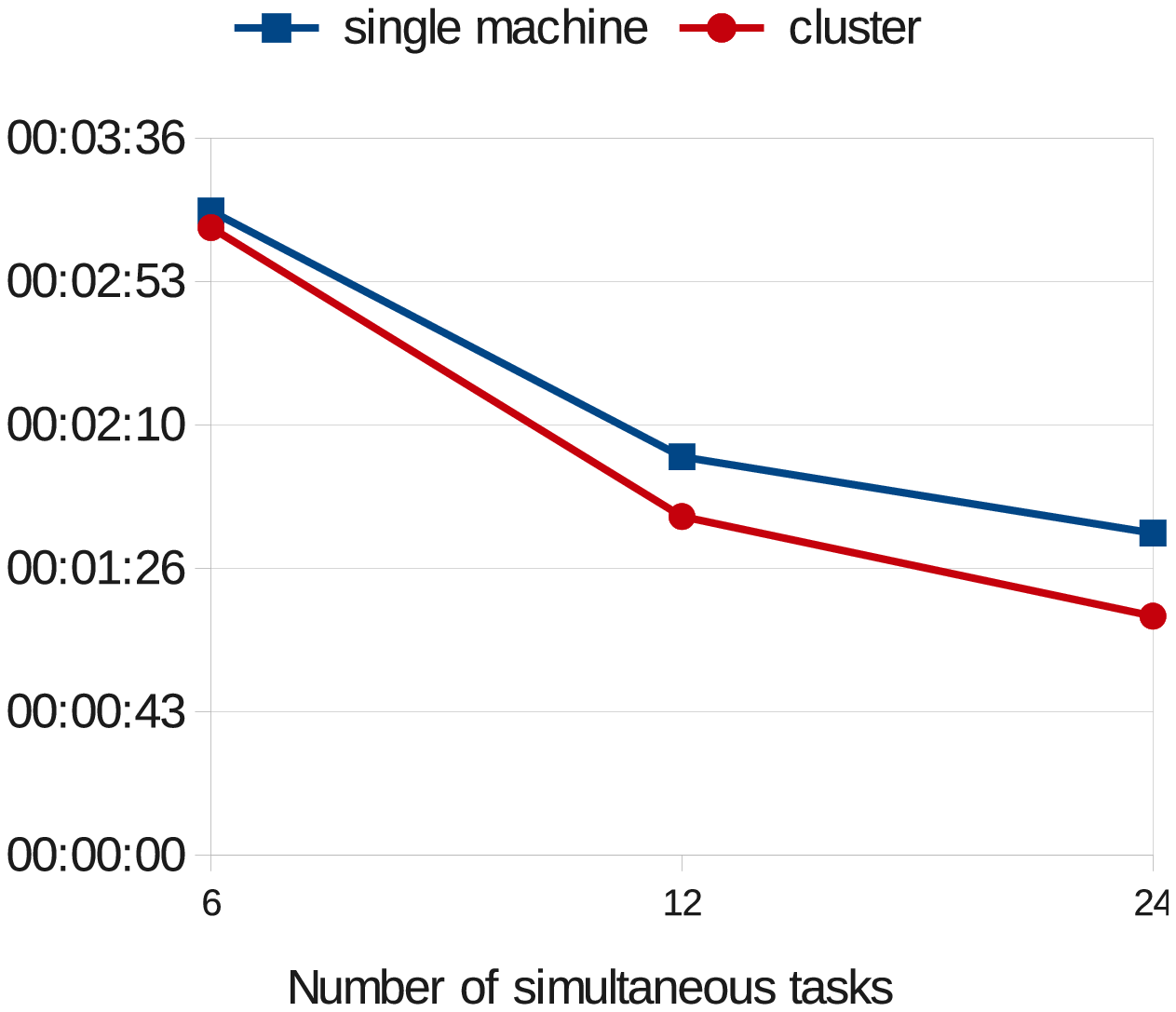}}
\subfloat[10,000 images]{\label{fig:10k}\includegraphics[scale=0.4]{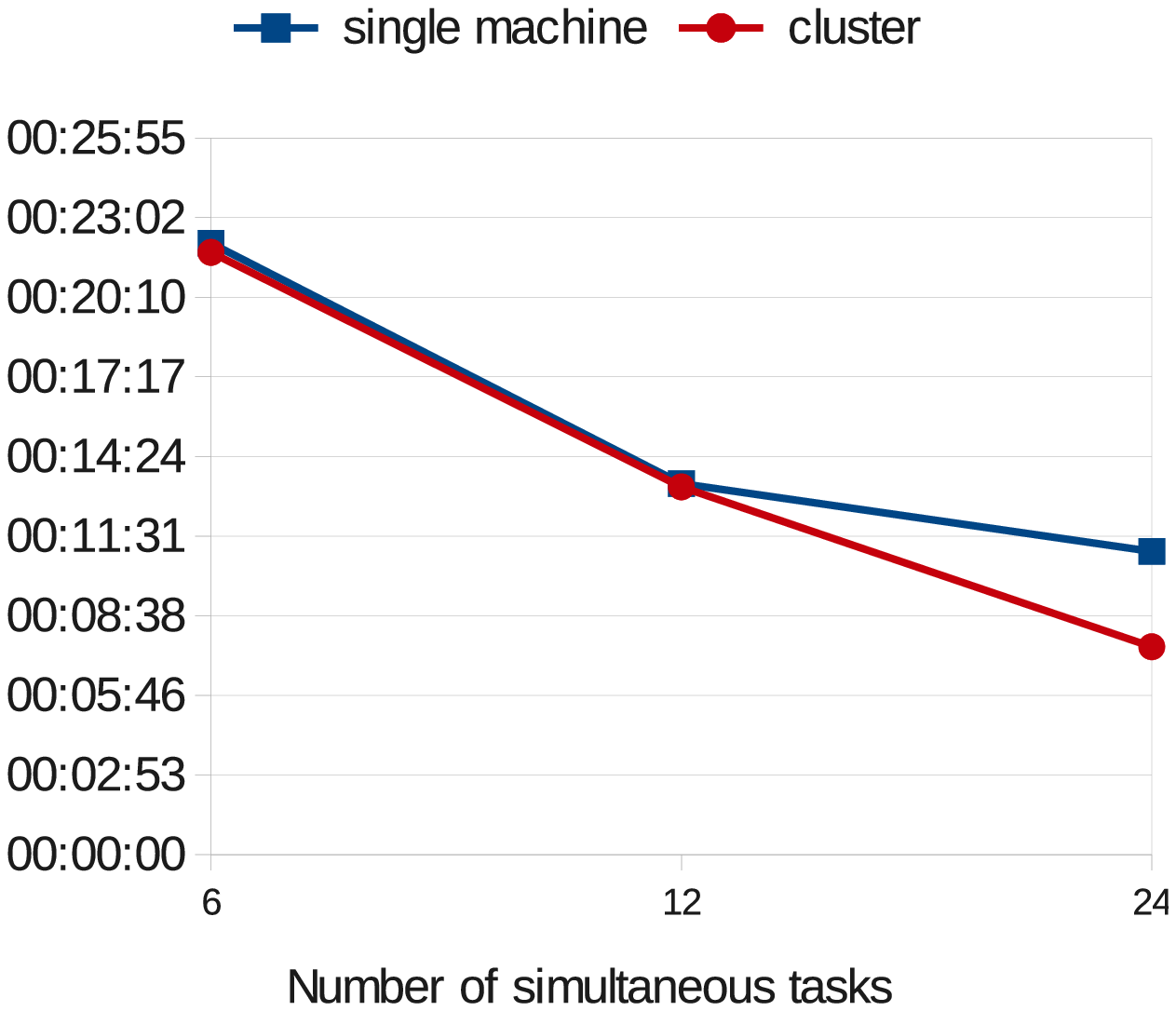}}
\subfloat[100,000 images]{\label{fig:100k}\includegraphics[scale=0.4]{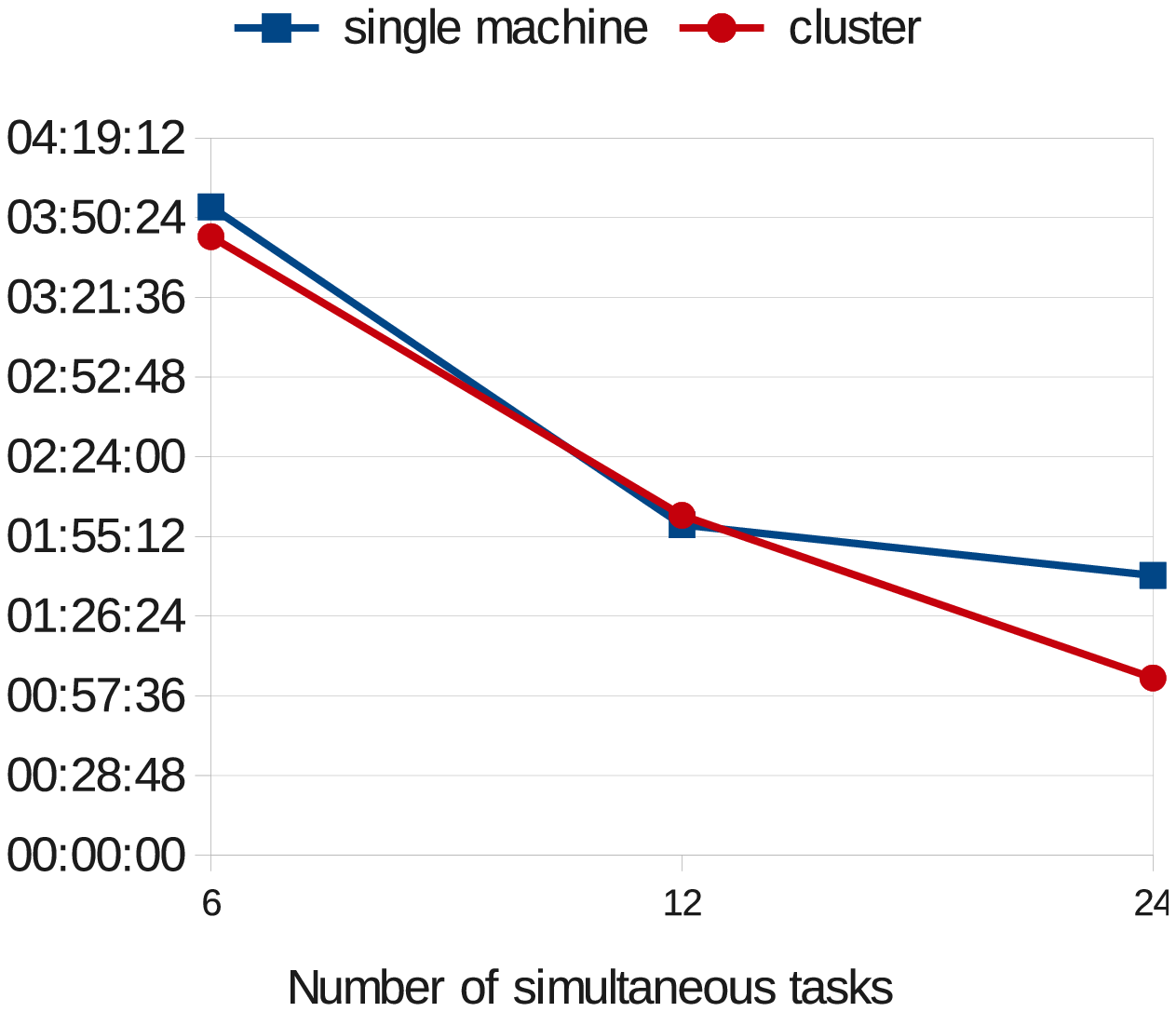}}
\caption{Comparison of runtimes using a single node and the distributed Hadoop cluster. 
For the datasets of 10k and 100k images, the single machine run does not scale linearly with more than 12 simultaneous tasks.}\label{fig:indexing}
\end{figure*}

For the monolithic runs 2), two approaches were taken. 
The first one uses 6, 12 and 24 simultaneous tasks on a single node (i.e., the PowerEdge R710 Server) for the indexing MapReduce run, 
while the second uses the same number of tasks distributed among the nodes of the Hadoop cluster. 
A text file containing a list of the image file paths was given as input and the same number (50) of images per map task, was given in all runs. 
A single reducer was used in all runs. The results are shown in Figure~\ref{fig:indexing}. 

%% file: parts/4_4_results_riesz3d.tex
\subsection{Riesz 3D}
The particularity of this use case resides in the fact that the available application was developed as a series of Matlab\textsuperscript{\textregistered} scripts (a proprietary language for the Matlab\textsuperscript{\textregistered} numerical computing environment\footnote{\texttt{http://www.mathworks.com/products/matlab/}, as of 29 August 2012.}), 
and a considerable effort would have been required to translate the whole code into Java, the native Hadoop programming language.
Luckily, a Hadoop feature called streaming allows using any executable written in any programming language to be used as the mapper and reducer of a Hadoop job. 
The only requirement for Hadoop streaming is that the executables must be able to read from the standard input stream \texttt{stdin} and write to the standard output stream \texttt{stdout}\footnote{\texttt{http://hadoop.apache.org/common/docs/r0.15.2/\\streaming.html}, as of 29 August 2012.}.
Using this feature, it was possible to keep to majority of the code as--is, requiring only minor adaptations to create a Mapper and Reducer script.
In order to maximize the scalability of this solution and avoid monopolizing several Matlab\textsuperscript{\textregistered} licenses for extended periods of time, the free and open--source Matlab\textsuperscript{\textregistered} clone GNU Octave\footnote{\texttt{http://www.gnu.org/software/octave/}, as of 29 August 2012.} was deployed on all the nodes of the cluster as a runtime environment for the scripts, which ran successfully without any modifications.

The runtime of the Riesz 3D use case with $s=4$, $N=4$, $d=3$ in Eq.~(\ref{eq:rieszComp}) was measured in the following three scenarios: 
one single task on one host, without using Hadoop; 
42 concurrent tasks on the Hadoop cluster when the nodes were idle; 
42 concurrent tasks on the Hadoop cluster with a normal load on the nodes. 
The respective runtimes obtained are 131h32m42s, 6h29m51s and 5h51m31s.
%

%% file: parts/5_discussion.tex
\section{Discussions and Conclusions}
Large--scale medical image analysis and retrieval is addressed in this work using the MapReduce framework. 
Three medical image analysis use--cases were implemented: 
optimization of SVMs for lung image classification, 
image indexing for content--based medical image search, 
and three--dimensional texture processing. 
The three use--cases reflect the various challenges of processing medical visual information in clinical routine: 
parameter optimization, indexation of image collections with hundreds of thousands images, and multi--dimensional medical data processing.

A cluster of heterogeneous computing nodes was set up in our institution using Hadoop, 
which currently allows for a maximum of 42 concurrent map tasks on 9 distinct machines, from which 8 are simple desktop computers with daily users. 
The cluster was designed to be minimally invasive on each machine and regular users of the desktop computers did not encounter problems when Hadoop jobs were running. 
The runtime variability is depicted in Fig.~\ref{fig:JobVariability}, which shows an excellent job runtime stability. 
This stability is the result of the cluster design, which does not allow for overbooking the computing resources. 
Jobs that were run between 10:00 and 12:00 as well as 18:00 and 22:00 generally ran faster than the ones executed during the night or the afternoon. 
The same occurs for the weekday versus week--end differentiation, which shows that both categories span the whole spectrum of measured execution times. 
A maximum difference of 28\% was observed between the shortest time (7m) and the longest one (8m58s). This variation is expected to be smaller with long--running, 
computationally intensive jobs, for which the overall impact of a temporarily slowed down CPU or network connection should be less significant.

Parallel grid search for optimal SVM parameters was carried out on the Hadoop cluster. 
The suspected link between the runtime of a map task $(C,\sigma)$ required for solving Eq.~(\ref{eq:SVM}) 
and the resulting classification accuracy was confirmed and is well illustrated in Fig.~\ref{fig:SVMOVAGraph}. 
It can be observed that most of the tasks with long runtimes resulted in poor classification accuracies 
and that no $(C,\sigma)$ value couple leading to best accuracies was interrupted. 
The total runtime was reduced from 50h to 9h15m using the task interruption method.
A minimum number of concurrent map tasks is required to ensure that some fastest tasks are included in the initial estimation of the runtime. 

Two approaches for content--based image indexing were compared and implemented in the MapReduce framework: component--based versus monolithic indexing. 
The former is convenient to separately optimize feature extraction and the bag--of--visual--words indexer 
because it does not require to run the whole pipeline for each optimization. 
However this costs time and storage space when you perform the whole pipeline due to the output writes and reads. 
This was observed with an unexpectedly long runtime for the feature extractor with the MapReduce framework in the component--based approach. 
The process was slowed down because it requires to write the features to a very large CSV (Comma--Separated Values) file of approximately 100 Gb for 100,000 images. 
The result is consistent with previous work that showed that the MapReduce framework was not performing well with IO--intensive tasks~\cite{LNK2009}.
The monolithic strategy showed to be well--suited to the MapReduce framework, which allowed indexing 100,000 images in about one hour using 24 concurrent tasks (see Fig.~\ref{fig:indexing} (c)).
The performances of the monolithic approach when running on a single node versus distributed nodes is compared. 
Whereas runtimes using 6 concurrent map tasks are similar (see Fig.~\ref{fig:indexing} (a) and (b)), in the case of 12 and 24 nodes the results for large datasets differ. 
In contrast to the multi--node approach, the single--node times do not scale linearly. 
This illustrates the advantage of using several nodes, which benefits from the distributed memory and disk access of each separated node.

\begin{figure}[h]
\centering
\includegraphics[scale=0.285,trim = 0mm 40mm 0mm 0mm]{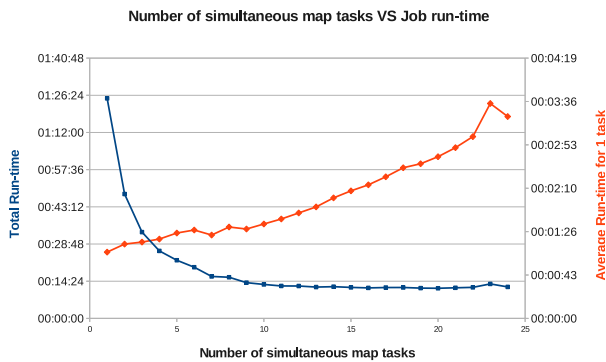}
\caption{Evolution of the job's total runtime and average runtime of a single task with an increasing number of simultaneous tasks executed on one host for the Riesz 3D use case in Octave.}
\label{fig:OctaveGraph}
\end{figure}

The parallelization of solid texture processing based on non--separable three--dimensional Riesz wavelets allowed to reduce a total runtime from more than 130h to less than 6h, 
while keeping the code in the original Matlab/Octave programming language with Hadoop streaming. 
There is a 22x speedup between the single--task sequential execution and the fastest Hadoop job. 
It is also noteworthy to point out that the execution during an idle period (during which all the nodes including the powerful server were not used for any other intensive tasks) actually performed worse than the job that was executed in more standard conditions (with no special precautions taken to free up resources on the nodes). This fact confirms that there is a non--deterministic aspect to the execution of Hadoop jobs. 
Using Hadoop streaming saves time and hassle in the development phase, but it also has a few drawbacks. 
First of all, a performance loss is inevitable when using the streaming feature as opposed to a ``native" Java program. 
Second, the existing application (e.g., Octave) may not be optimized for parallel usage, resulting in a non--negligible slowdown when using a large number of simultaneous tasks. 
Figure~\ref{fig:OctaveGraph} shows that the runtime gain decreases rapidly as the average runtime for a single task increases in a fairly linear fashion. Hardware bottlenecks such as shared CPU cache memory or non--optimal utilization of Intel\textsuperscript{\textregistered}'s Hyper--threading\footnote{\texttt{http://www.intel.com/content/www/us/en/\\architecture-and-technology/hyper-threading/\\hyper-threading-technology.html}, as of 29 August 2012.} feature are a probable cause of this slowdown. 
These results are aligned with the observations on the non--linear scaling of runtimes for the monolithic image indexing use case in Figure~\ref{fig:indexing} (c)
and highlights once more the advantage of running concurrent map tasks on separated nodes using Hadoop.

In future work, we plan to extend the cluster on demand using cloud computing such as Amazon Elastic MapReduce (Amazon EMR)\footnote{\texttt{http://aws.amazon.com/elasticmapreduce/}, as of 29 August 2012.}.
This would allow to mix local resources with fully manage resources that are potentially available in large number and can help to increase response times at  limited cost.

Overall Hadoop has shown its utility for large scale medical image computing. In most tasks very positive results could be obtained helping the projects to scale with limited local resources available.